\documentclass{article}
\usepackage{amssymb}

\usepackage{graphicx}
\usepackage{amsmath}

\input{tcilatex}

\begin{document}

\title{\textsf{An index }$_{2}F_{2}$ \textsf{hypergeometric transform }}
\author{\textsf{Zouha\"{i}r MOUAYN}}
\date{{\small Sultan Moulay Slimane University, Faculty of \ Sciences and Technics
(M'Ghila) PO.Box 523, B\'{e}ni Mellal, Morocco }\\
{\small mouayn@gmail.com}}
\maketitle

\begin{abstract}
We construct a new one-parameter family of index hypergeometric transforms
associated with the relativistic pseudoharmonic oscillator by using coherent
states analysis.
\end{abstract}

\section{Introduction}

In $\left[ 1\right] ,$ while introducing a class of coherent states attached
to symmetric spaces of non-compact type, the authors have considered a
family of weighted Bergman spaces labelled by a real parameter $\gamma $
with $2\gamma =1,2,...$, as 
\begin{equation}
\frak{F}_{\gamma }\left( \mathbb{C}\right) =\left\{ \psi \text{ analytic on }%
\mathbb{C}\text{,}\int\limits_{\mathbb{C}}\left| \psi \left( z\right)
\right| ^{2}K_{\frac{1}{2}-\gamma }\left( 2\left| z\right| \right) \left|
z\right| ^{2\gamma -1}d\mu \left( z\right) <\infty \right\}  \tag{1.1}
\end{equation}
where $K_{\nu }\left( .\right) $ denotes the MacDonald function $\left[ 2%
\right] $ and $d\mu \left( z\right) $ is the Lebesgue measure on $\mathbb{C}%
. $

In this paper, we construct a one-parameter family of index hypergeometric
transforms\ mapping isometrically the space of complex-valued square
integrable functions $\varphi \left( x\right) $ on positive real half-line
onto spaces in $\left( 1.1\right) $ by 
\begin{equation*}
\digamma _{\gamma }:L^{2}\left( \mathbb{R}_{+},dx\right) \rightarrow \frak{F}%
_{\gamma }\left( \mathbb{C}\right) 
\end{equation*}
\begin{equation}
\digamma _{\gamma }\left[ \varphi \right] \left( z\right) =\frac{i^{\gamma
}\exp \left( z\right) }{\Gamma \left( 2\gamma \right) \Gamma \left( \gamma +%
\frac{1}{2}\right) }\int\limits_{0}^{+\infty }\frac{_{2}\digamma _{2}\left( 
\begin{array}{c}
\gamma +ix,\gamma -ix \\ 
2\gamma ,\gamma +\frac{1}{2}
\end{array}
;z\right) }{\left( 2\gamma \left( \gamma -1\right) \right) ^{ix}\Gamma
\left( ix\right) \Gamma ^{-2}\left( \gamma +ix\right) }\varphi \left(
x\right) dx  \tag{1.2}
\end{equation}
with the Euler gamma function $\Gamma $ and the $_{2}F_{2}$ hypergeometric
series $\left[ 3\right] .$

Our method is based on coherent state analysis. That is, we present a class
of coherent states as superpositions of eigenstates of a relativistic model
for the pseudoharmonic oscillator $\left[ 4\right] .$ These eigenstates are
expressed in terms of continuous dual Hahn polynomials $\left[ 3\right] $.
In this superposition, the coefficients are the basis elements of the
weighted Bergman spaces in $\left( 1.1\right) $. The closed form of the
constructed coherent states allows to define a coherent states transform
which turns out to be the index hypergeometric transform in $\left(
1.2\right) .$

The paper is organized as follows. Section 2 deals with a brief formalism of
coherent states and their corresponding coherent state transforms. In
section 3, we summarize some needed tools on the relativistic pseudoharmonic
oscillator. In section 4, we recall the definition of the weighted Bergman
spaces we are dealing with. In section 5, we construct a class of coherent
states from which we deduce a one-parameter family of index hypergeometric
transforms.

\section{Coherent states}

Following $\left[ 5\right] $, let $(X,\sigma )$\ be a measure space and let $%
\mathcal{A}_{2}\left( X\right) \subset L^{2}(X,\sigma )$\ be a closed
subspace of infinite dimension. Let $\left\{ \Phi _{n}\right\}
_{n=0}^{\infty }$ be an orthogonal basis of $\mathcal{A}_{2}\left( X\right) $
satisfying, for arbitrary $\xi \in X,$

\begin{equation}
\omega \left( \xi \right) :=\sum_{n=0}^{\infty }\frac{\left| \Phi _{n}\left(
\xi \right) \right| ^{2}}{\rho _{n}}<+\infty ,  \tag{2.1}
\end{equation}
where $\rho _{n}:=\left\| \Phi _{n}\right\| _{L^{2}(X)}^{2}$\ . Define

\begin{equation}
\mathcal{K}(\xi ,\zeta ):=\sum_{n=0}^{\infty }\frac{\Phi _{n}\left( \xi
\right) \overline{\Phi _{n}(\zeta )}}{\rho _{n}},\text{ }\xi ,\zeta \in X. 
\tag{2.2}
\end{equation}
Then, $\mathcal{K}(\xi ,\zeta )$\ is a reproducing kernel, $\mathcal{A}%
_{2}\left( X\right) $ is the corresponding reproducing kernel Hilbert space
and $\omega \left( \xi \right) =\mathcal{K}(\xi ,\xi )$, $\xi \in X.$

Let $\mathcal{H}$ be another Hilbert space with $\dim \mathcal{H}=\infty $
and $\left\{ \phi _{n}\right\} _{n=0}^{\infty }$\ be an orthonormal basis of 
$\mathcal{H}.$ Therefore$,$ define a coherent state as a ket vector $\mid
\xi >\in \mathcal{H}$ labelled by a point $\xi \in X$ as 
\begin{equation}
\mid \xi >:=\left( \omega \left( \xi \right) \right) ^{-\frac{1}{2}%
}\sum_{n=0}^{\infty }\frac{\Phi _{n}\left( \xi \right) }{\sqrt{\rho _{n}}}%
\mid \phi _{n}>.  \tag{2.3}
\end{equation}
We rewrite $\left( 2.3\right) $ using Dirac's bra-ket notation as 
\begin{equation}
<x\mid \xi >=\left( \omega \left( \xi \right) \right) ^{-\frac{1}{2}%
}\sum_{n=0}^{\infty }\frac{\Phi _{n}\left( \xi \right) }{\sqrt{\rho _{n}}}%
<x\mid \phi _{n}>.  \tag{2.4}
\end{equation}
By definition, it is straightforward to show that $<\xi \mid \xi >=1$\ and
the coherent state transform $W:\mathcal{H\rightarrow A}_{2}\left( X\right)
\subset L^{2}(X,\sigma )$ defined by 
\begin{equation}
W\left[ \phi \right] \left( \xi \right) :=\left( \omega \left( \xi \right)
\right) ^{\frac{1}{2}}<\xi \mid \phi >  \tag{2.5}
\end{equation}
is an isometry. Thus, for $\phi ,\psi \in \mathcal{H}$, we have 
\begin{equation}
<\phi \mid \psi >_{\mathcal{H}}=<W\left[ \phi \right] \mid W\left[ \psi %
\right] >_{L^{2}\left( X\right) }=\int\limits_{X}d\sigma \left( \xi \right)
\omega \left( \xi \right) <\phi \mid \xi ><\xi \mid \psi >  \tag{2.6}
\end{equation}
and thereby we have a resolution of the identity 
\begin{equation}
\mathbf{1}_{\mathcal{H}}=\int\limits_{X}d\sigma \left( \xi \right) \omega
\left( \xi \right) \mid \xi ><\xi \mid ,  \tag{2.7}
\end{equation}
where $\omega \left( \xi \right) $\ appears as a weight function.

\section{The weighted Bergman space $\frak{F}_{\protect\gamma }\left( 
\mathbb{C}\right) $}

As mentioned in Section.1, a countable set of Hilbert spaces whose elements
are analytic functions on $\mathbb{C},$ have been introduced by the authors
in ($\left[ 1\right] $, p.51) as follows. For each fixed $\gamma $ with $%
2\gamma =1,2,...$ $,$ the inner product is defined by 
\begin{equation}
\left\langle \psi ,\phi \right\rangle _{\gamma }:=\int\limits_{\mathbb{C}%
}\psi \left( z\right) \overline{\phi \left( z\right) }d\mu _{\gamma }\left(
z\right) ,  \tag{3.1}
\end{equation}
where 
\begin{equation}
d\mu _{\gamma }\left( z\right) =\frac{2}{\pi \Gamma \left( 2\gamma \right) }%
\rho ^{2\gamma -1}K_{\frac{1}{2}-\gamma }\left( 2\rho \right) \rho d\theta
d\rho ,z=\rho e^{i\theta }\in \mathbb{C}  \tag{3.2}
\end{equation}
with the MacDonald function defined by ($\left[ 5\right] $, p.183): 
\begin{equation}
K_{\nu }\left( \rho \right) =\frac{1}{2}\left( \frac{\rho }{2}\right) ^{\nu
}\int\limits_{0}^{+\infty }t^{-\nu -1}\exp \left( -t-\frac{\rho ^{2}}{4t}%
\right) dt.  \tag{3.3}
\end{equation}
Using the notation in $\left( 3.1\right) $, the space in $\left( 1.1\right) $
also reads 
\begin{equation}
\frak{F}_{\gamma }\left( \mathbb{C}\right) =\left\{ \psi \text{ analytic on }%
\mathbb{C}\text{, }\left\langle \psi ,\psi \right\rangle _{\gamma }<+\infty .%
\text{\ }\right\}   \tag{3.4}
\end{equation}
A well known orthonormal basis of the space in $\left( 3.4\right) $ is given
by the functions 
\begin{equation}
\psi _{n}^{\gamma }(z):=\frac{z^{n}}{\sqrt{n!\left( 2\gamma \right) _{n}}},%
\text{ }n=0,1,2,...,\text{ }z\in \mathbb{C},  \tag{3.5}
\end{equation}
and its reproducing kernel is obtained as  
\begin{equation}
\mathcal{K}_{\gamma }\left( z,w\right) :=\sum\limits_{n=0}^{+\infty }\frac{1%
}{\left( 2\gamma \right) _{n}}\frac{\left( z\overline{w}\right) ^{n}}{n!} 
\tag{3.6}
\end{equation}
in which $\left( a\right) _{n}$ is the Pochhammer symbol defined by $\left(
a\right) _{0}:=1$ and 
\begin{equation}
\left( a\right) _{n}:=a\left( a+1\right) ...\left( a+n-1\right) =\frac{%
\Gamma \left( a+n\right) }{\Gamma \left( a\right) }.  \tag{3.7}
\end{equation}
One can use the formula for the modified Bessel function of the first kind\ (%
$\left[ 5\right] ,$ p.77): 
\begin{equation}
I_{\nu }\left( \zeta \right) =\sum\limits_{n=0}^{+\infty }\frac{1}{n!\Gamma
\left( \nu +n+1\right) }\left( \frac{\zeta }{2}\right) ^{\nu +2n}\text{ ,} 
\tag{3.8}
\end{equation}
for $\zeta =2\sqrt{z\overline{w}}$ and $\nu =2\gamma -1,$ to present $\left(
3.6\right) $ in a closed form as 
\begin{equation}
\mathcal{K}_{\gamma }\left( z,w\right) =\Gamma \left( 2\gamma \right) \left(
z\overline{w}\right) ^{\frac{1}{2}-\gamma }I_{2\gamma -1}\left( 2\sqrt{z%
\overline{w}}\right) ,  \tag{3.9}
\end{equation}
So that the diagonal function of $\left( 3.9\right) $ reads  
\begin{equation}
\mathcal{K}_{\gamma }\left( z,z\right) =\Gamma \left( 2\gamma \right) \left|
z\right| ^{1-2\gamma }I_{2\gamma -1}\left( 2\left| z\right| \right) ,z\in 
\mathbb{C}.  \tag{3.10}
\end{equation}
The latter will be used in the sequel.

\section{A relativistic model for the pseudoharmonic oscillator}

In this section, we recall some needed results which have been developed in $%
\left[ 4\right] $ where the authors considered a model for the relativistic
pseudoharmonic oscillator with the following interaction potential 
\begin{equation}
V\left( x\right) :=\left( \frac{1}{2}m\omega ^{2}x\left( x+i\lambda \right) +%
\frac{g}{x\left( x+i\lambda \right) }\right) e^{i\lambda \partial _{x}} 
\tag{4.1}
\end{equation}
where $\omega $ is a frequency, $g\geq 0$ is a real quantity and $\lambda
=\hbar /mc$ denotes the Compton wavelength defined by the ratio of Planck's
constant $\hbar $ by the mass $m$ times the speed of light $c.$

The corresponding stationary Schr\"{o}dinger equation is described by the
finite-difference equation 
\begin{equation}
\left( mc^{2}\cosh i\lambda \partial _{x}+\frac{1}{2}m\omega ^{2}x\left(
x+i\lambda \right) e^{i\lambda \partial _{x}}+\frac{g}{x\left( x+i\lambda
\right) }e^{i\lambda \partial _{x}}\right) \varphi \left( x\right) =\epsilon
\varphi \left( x\right)   \tag{4.2}
\end{equation}
with the boundary conditions for the wave function $\varphi \left( 0\right)
=0$ and $\varphi \left( \infty \right) =0$. The energy spectrum of the Schr%
\"{o}dinger operator in (4.2) is 
\begin{equation}
\epsilon _{n}:=\hbar \omega \left( 2n+\alpha _{+}+\alpha _{-}\right)
,n=0,1,2,...  \tag{4.3}
\end{equation}
where 
\begin{equation}
\alpha _{\pm }=\frac{1}{2}+\frac{1}{2}\sqrt{1+\frac{2}{\omega _{0}}\left(
1\pm \sqrt{1-8g_{0}\omega _{0}^{2}}\right) }  \tag{4.4}
\end{equation}
and 
\begin{equation}
\omega _{0}=\frac{\hbar \omega }{mc^{2}}\text{ , \ \ \ \ \ \ }g_{0}=\frac{mg%
}{\hbar ^{2}}  \tag{4.5}
\end{equation}
The corresponding orthonormalized eigenstates of the form 
\begin{equation}
<x\mid \varphi _{n}^{\alpha _{\pm }}>=c_{n}\left( -x\right) ^{(\alpha
_{+})}\omega _{0}^{ix}\Gamma \left( \alpha _{-}+ix\right) S_{n}\left(
x^{2};\alpha _{+},\alpha _{-},\frac{1}{2}\right)   \tag{4.6}
\end{equation}
where 
\begin{equation}
c_{n}=\left( 2^{-1}n!\Gamma \left( n+\alpha _{+}+\alpha _{-}\right) \Gamma
\left( n+\alpha _{+}+\frac{1}{2}\right) \Gamma \left( n+\alpha _{-}+\frac{1}{%
2}\right) \right) ^{-\frac{1}{2}}  \tag{4.7}
\end{equation}
and 
\begin{equation}
S_{n}\left( x^{2};a,b,c\right) :=\left( a+b\right) _{n}\left( a+c\right)
_{n}{}_{3}F_{2}\left( 
\begin{array}{c}
-n,a+ix,a-ix \\ 
a+b,a+c
\end{array}
;1\right)   \tag{4.8}
\end{equation}
is the continuous dual Hahn polynomial ($\left[ 3\right] $, p.331) given in
terms of the ${}_{3}F_{2}$-sum. The notation 
\begin{equation}
\zeta ^{\left( \nu \right) }:=i^{\nu }\frac{\Gamma \left( \nu -i\zeta
\right) }{\Gamma \left( -i\zeta \right) }  \tag{4.9}
\end{equation}
means the generalized degree ($\left[ 4\right] $,$\left[ 6\right] $)$.$ The
wavefunctions in $\left( 4.6\right) $ satisfy the relations $:$ 
\begin{equation}
\int\limits_{0}^{+\infty }<x\mid \varphi _{m}^{\alpha _{\pm }}>\overline{%
<x\mid \varphi _{n}^{\alpha _{\pm }}>}dx=\delta _{n,m}  \tag{4.10}
\end{equation}
which means that they constitute an orthonormalized system in the Hilbert
space $L^{2}\left( \mathbb{R}_{+},dx\right) .$

\section{An index hypergeometric transform}

We now combine the two basis $\left( \psi _{n}^{\gamma }\right) _{n}$ in $%
\left( 3.5\right) $ and $\left( \mid \varphi _{n}^{\alpha _{\pm }}>\right)
_{n}$ in $\left( 4.6\right) $ together with the function $\mathcal{K}%
_{\gamma }\left( .,.\right) $ in $\left( 3.10\right) $ to construct for
every fixed $\gamma \in \mathbb{R}$ with $2\gamma =1,2,...,$ a set of
coherent states $\left( \mid z;\gamma >\right) _{z\in \mathbb{C}}$ labelled
by points $z$ $\in \mathbb{C}$ according to definition $\left( 2.3\right) $
by setting 
\begin{equation}
\mid z;\gamma >:=\left( \mathcal{K}_{\gamma }\left( z,z\right) \right) ^{-%
\frac{1}{2}}\sum_{n=0}^{\infty }\frac{z^{n}}{\sqrt{\left( 2\gamma \right)
_{n}n!}}\mid \varphi _{n}^{\alpha _{\pm }}>  \tag{5.1}
\end{equation}
For our purpose, we choose the parameters $g$ and $\omega $ in $\left(
4.1\right) $ are such that $8g\omega ^{2}=mc^{4}.$ In such case and in view
of $\left( 4.5\right) ,$ we get that 
\begin{equation}
1-8g_{0}\omega _{0}^{2}=0  \tag{5.2}
\end{equation}
and Eq.$\left( 4.4\right) $ reduces to 
\begin{equation}
\alpha _{+}=\alpha _{-}=\frac{1}{2}+\frac{1}{2}\sqrt{1+\frac{2}{\omega _{0}}}
\tag{5.3}
\end{equation}
Now, if we set 
\begin{equation}
\gamma :=\alpha _{+}=\alpha _{-}  \tag{5.4}
\end{equation}
which means that 
\begin{equation}
2\gamma -1=\sqrt{1+\frac{16g\omega }{\hbar c^{2}}}=\sqrt{1+2\frac{mc^{2}}{%
\hbar \omega }}.  \tag{5.5}
\end{equation}
Then, we can establish a closed form for wavefunctions of the coherent
states defined in $\left( 5.1\right) $ as follows. Let $x\in \mathbb{R}_{+}.$
Then, starting from $\left( 5.1\right) $ with the condition $\left(
5.4\right) ,$ we can write successively 
\begin{equation}
<x\mid z;\gamma >=\left( \mathcal{K}_{\gamma }\left( z,z\right) \right) ^{-%
\frac{1}{2}}\sum_{n=0}^{\infty }\frac{z^{n}}{\sqrt{\left( 2\gamma \right)
_{n}n!}}<x\mid n;\gamma >,  \tag{5.6}
\end{equation}
\begin{equation}
=\left( \mathcal{K}_{\gamma }\left( z,z\right) \right) ^{-\frac{1}{2}%
}\sum_{n=0}^{\infty }\frac{c_{n}z^{n}}{\sqrt{\left( 2\gamma \right) _{n}n!}}%
\left( -x\right) ^{(\gamma )}\omega _{0}^{ix}\Gamma \left( \gamma +ix\right)
S_{n}\left( x^{2};\gamma ,\gamma ;\frac{1}{2}\right)   \tag{5.7}
\end{equation}
\begin{equation}
=\frac{\left( -x\right) ^{(\gamma )}\omega _{0}^{ix}\Gamma \left( \gamma
+ix\right) }{\sqrt{\mathcal{K}_{\gamma }\left( z,z\right) }}%
\sum_{n=0}^{\infty }\frac{z^{n}}{\sqrt{\left( 2\gamma \right) _{n}n!}}%
c_{n}S_{n}\left( x^{2};\gamma ,\gamma ,\frac{1}{2}\right)   \tag{5.8}
\end{equation}
\begin{equation}
=\frac{\left( -x\right) ^{(\gamma )}\omega _{0}^{ix}\Gamma \left( \gamma
+ix\right) }{\sqrt{\mathcal{K}_{\gamma }\left( z,z\right) }}%
\sum_{n=0}^{\infty }\frac{S_{n}\left( x^{2};\gamma ,\gamma ,\frac{1}{2}%
\right) }{\sqrt{\left( 2\gamma \right) _{n}}\sqrt{\Gamma \left( 2\gamma
+n\right) }\Gamma \left( n+\gamma +\frac{1}{2}\right) }\frac{z^{n}}{n!} 
\tag{5.9}
\end{equation}
Let us set 
\begin{equation}
<x\mid z;\gamma >=\frac{\left( -x\right) ^{(\alpha )}\omega _{0}^{ix}\Gamma
\left( \gamma +ix\right) }{\sqrt{\omega _{\gamma }\left( z\right) }}\frak{G}%
_{\gamma }\left( x,z\right)   \tag{5.10}
\end{equation}
where 
\begin{equation}
\frak{G}_{\gamma }\left( x,z\right) :=\sum_{n=0}^{\infty }\frac{S_{n}\left(
x^{2};\gamma ,\gamma ,\frac{1}{2}\right) }{\sqrt{\left( 2\gamma \right) _{n}}%
\sqrt{\Gamma \left( 2\gamma +n\right) }\Gamma \left( n+\gamma +\frac{1}{2}%
\right) }\frac{z^{n}}{n!}  \tag{5.11}
\end{equation}
\begin{equation}
=\frac{1}{\sqrt{\Gamma \left( 2\gamma \right) }\Gamma \left( \gamma +\frac{1%
}{2}\right) }\sum_{n=0}^{\infty }\frac{S_{n}\left( x^{2};\gamma ,\gamma ,%
\frac{1}{2}\right) }{\left( 2\gamma \right) _{n}\left( \gamma +\frac{1}{2}%
\right) _{n}}\frac{z^{n}}{n!}  \tag{5.12}
\end{equation}
Making use of the following generating formula for the continuous dual Hahn (%
$\left[ 3\right] ,$p.349): 
\begin{equation}
e^{\xi }{}_{2}\digamma _{2}\left( 
\begin{array}{c}
a+ix,a-ix \\ 
a+b,a+c
\end{array}
;-\xi \right) =\sum\limits_{n=0}^{+\infty }\frac{S_{n}\left(
x^{2};a,b,c\right) }{\left( a+b\right) _{n}\left( a+c\right) _{n}}\frac{\xi
^{n}}{n!}  \tag{5.13}
\end{equation}
for $a=b=\gamma $, $c=\frac{1}{2}$ and $\xi =z,$ we obtain that 
\begin{equation}
\frak{G}_{\gamma }\left( x,z\right) =\frac{e^{z}{}}{\sqrt{\Gamma \left(
2\gamma \right) }\Gamma \left( \gamma +\frac{1}{2}\right) }._{2}\digamma
_{2}\left( 
\begin{array}{c}
\gamma +ix,\gamma -ix \\ 
2\gamma ,\gamma +\frac{1}{2}
\end{array}
;-z\right) .  \tag{5.14}
\end{equation}
Returning back to $\left( 5.10\right) $ and inserting $\left( 5.14\right) $,
we get that 
\begin{equation}
<x\mid z;\gamma >=\frac{\left( -x\right) ^{(\alpha )}\omega _{0}^{ix}\Gamma
\left( \gamma +ix\right) e^{z}{}}{\sqrt{\mathcal{K}_{\gamma }\left(
z,z\right) }}\frac{_{2}\digamma _{2}\left( 
\begin{array}{c}
\gamma +ix,\gamma -ix \\ 
2\gamma ,\gamma +\frac{1}{2}
\end{array}
;-z\right) }{\sqrt{\Gamma \left( 2\gamma \right) }\Gamma \left( \gamma +%
\frac{1}{2}\right) }  \tag{5.15}
\end{equation}
Next, making use of $\left( 4.9\right) ,$ we write $\left( -x\right)
^{(\gamma )}$ in $\left( 5.15\right) $ as 
\begin{equation}
\left( -x\right) ^{(\gamma )}=i^{\gamma }\frac{\Gamma \left( \gamma
+ix\right) }{\Gamma \left( ix\right) }  \tag{5.16}
\end{equation}
On the other hand, by $\left( 5.3\right) $ and $\left( 5.4\right) ,$ we can
write $\omega _{0}$ in terms of $\gamma $ as 
\begin{equation}
\omega _{0}=\frac{1}{\gamma \left( 2\gamma -1\right) }  \tag{5.17}
\end{equation}
Now, summarizing up the above calculation , Eq.$\left( 5.15\right) $ takes
the form 
\begin{equation}
<x\mid z;\gamma >=\frac{i^{\gamma }\frac{\Gamma \left( \gamma +ix\right) }{%
\Gamma \left( ix\right) }\left( \frac{1}{2\gamma \left( \gamma -1\right) }%
\right) ^{ix}\Gamma \left( \gamma +ix\right) }{\sqrt{\mathcal{K}_{\gamma
}\left( z,z\right) }}\frac{e^{z}{}_{2}\digamma _{2}\left( 
\begin{array}{c}
\gamma +ix,\gamma -ix \\ 
2\gamma ,\gamma +\frac{1}{2}
\end{array}
;-z\right) }{\sqrt{\Gamma \left( 2\gamma \right) }\Gamma \left( \gamma +%
\frac{1}{2}\right) }  \tag{5.18}
\end{equation}
Finally, making use of Eq. $\left( 3.10\right) ,$ arrive at the following
closed form for 
\begin{equation}
<x\mid z;\gamma >=\frac{i^{\gamma }\left( 2\gamma \left( \gamma -1\right)
\right) ^{-ix}\Gamma ^{2}\left( \gamma +ix\right) e^{z}{}{}_{2}\digamma
_{2}\left( 
\begin{array}{c}
\gamma +ix,\gamma -ix \\ 
2\gamma ,\gamma +\frac{1}{2}
\end{array}
;-z\right) }{\sqrt{\left| z\right| ^{1-2\gamma }I_{2\gamma -1}\left( 2\left|
z\right| \right) }\Gamma \left( 2\gamma \right) \Gamma \left( \gamma +\frac{1%
}{2}\right) \Gamma \left( ix\right) }  \tag{5.19}
\end{equation}
Finally, by $\left( 2.5\right) ,$the coherent state transform corresponding
the coherent states\ in $\left( 5.6\right) $ with the closed form $\left(
5.19\right) $ of their wavefunctions is the isometry mapping 
\begin{equation}
\digamma _{\gamma }:L^{2}\left( \mathbb{R}^{+},dx\right) \rightarrow \frak{F}%
_{\gamma }\left( \mathbb{C}\right)   \tag{5.20}
\end{equation}
defined by 
\begin{equation}
\digamma _{\gamma }\left[ \varphi \right] \left( z\right) :=\left( \mathcal{K%
}_{\gamma }\left( z,z\right) \right) ^{\frac{1}{2}}<z;\gamma \mid \varphi
>_{L^{2}\left( \mathbb{R}^{+},dx\right) }  \tag{5.21}
\end{equation}
Explicitly, 
\begin{equation}
\digamma _{\gamma }\left[ \varphi \right] \left( z\right) =\frac{i^{\gamma
}\exp \left( z\right) }{\Gamma \left( 2\gamma \right) \Gamma \left( \gamma +%
\frac{1}{2}\right) }\int\limits_{0}^{+\infty }\frac{\Gamma ^{2}\left( \gamma
+ix\right) {}{}_{2}\digamma _{2}\left( 
\begin{array}{c}
\gamma +ix,\gamma -ix \\ 
2\gamma ,\gamma +\frac{1}{2}
\end{array}
;-z\right) }{\left( 2\gamma \left( \gamma -1\right) \right) ^{ix}\Gamma
\left( ix\right) }\overline{\varphi \left( x\right) }dx  \tag{5.22}
\end{equation}
which is the announced transform in $\left( 1.2\right) .$

\begin{center}
\textbf{References}
\end{center}

\begin{quote}
$\left[ 1\right] $ A. O. Barut and L. Girardello, \textit{Commun. Math. Phys}%
. \textbf{21}, 41 (1971)

$\left[ 2\right] $ G. N. Watson,\textit{\ Treatise on the Theory of Bessel
functions}, University Press,Cambridge 1958

$\left[ 3\right] $ G. E. Andrews, R. Askey and R. Roy, \textit{Special
Functions,} Encyclopedia of Mathematics and its Applications, Cambridge
University Press, 1999

$\left[ 4\right] $ S.M. Nagiyev, E.I. Jafarov and R. M. Imanov, \textit{J.
Phys. A: Math. Gen.}, \textbf{36 }(2003) p.7813

$\left[ 5\right] $ J.P. Gazeau, \textit{Coherent states in quantum physics},
Wiley-VCH Verlag GmbH \& KGaA Weinheim, 2009

$\left[ 6\right] $ M. Freeman, M. D. Mateev and R. M. Mir-Kasimov, \textit{%
Nucl. Phys. B} \textbf{12} (1969), p.197
\end{quote}

\end{document}